\documentstyle[psfig,12pt]{article}

\def\be{\begin{equation}}
\def\bea{\begin{eqnarray}}
\def\ee{\end{equation}}
\def\eea{\end{eqnarray}}

\begin{document}

\begin{titlepage}
\begin{flushright}
UMN-D-03-1\\
hep-th/0302119
\end{flushright}

\begin{centering}

\vspace*{3cm}

{\Large\bf Anomalously light mesons in a (1+1)-dimensional supersymmetric
theory with fundamental matter}

\vspace*{1.5cm}

{\bf J.R.~Hiller,$^a$ S.S.~Pinsky,$^b$ and U.~Trittmann$^b$\footnote{Present
address: Department of Physics, Otterbein College, Westerville OH 43081}}
\vspace*{0.5cm}

$^a$
{\sl Department of Physics \\
University of Minnesota Duluth\\
Duluth, MN  55812}

\vspace*{0.5cm}

$^b$
{\sl Department of Physics\\
Ohio State University\\
Columbus, OH 43210, USA}

\vspace*{1cm}


\vspace*{1cm}

{\large Abstract}

\vspace*{1cm}

\begin{abstract}
We consider ${\cal N}=1$ supersymmetric Yang--Mills theory with 
fundamental matter in the large-$N_c$ approximation in 1+1 dimensions. 
We add a Chern--Simons term to give the adjoint partons a mass and solve 
for the meson bound states. Here mesons are color-singlet states with two 
partons in the fundamental representation but are not necessarily bosons. 
We find that this theory has anomalously light meson bound states at 
intermediate and strong coupling.  We also examine the structure functions 
for these states and find that they prefer to have as many partons as 
possible at low longitudinal momentum fraction.
\end{abstract}
\end{centering}

\vfill

\end{titlepage}
\newpage
\section{Introduction.}

Supersymmetry is a property of some quantum field theories that provides a
beautiful solution to a host of theoretical problems~\cite{TPISUSY}. It is 
a pressing experimental issue to see if nature takes advantage of this elegant 
option. Of course we already know that supersymmetry is rather badly broken 
since we do not see any superpartners for the particles of the standard model. 
It is assumed that all of the superpartners are in fact very heavy and that 
we will see them as we go to higher energies in accelerators. The question 
then becomes, what are our expectations for the lightest superparticles that 
we might first see?  As we know from QCD, symmetries can give rise to very 
light bound states.  There are indications that the same thing can happen with
supersymmetry~\cite{Hiller:2002cu,Hiller:2002pj}.

Long ago 't Hooft~\cite{thooft} showed that two-dimensional models can be
powerful laboratories for the study of the bound-state problem. These models
remain popular to this day because they are easy to solve and share many
properties with their four-dimensional cousins, most notably stable bound 
states.  Supersymmetric two-dimensional models are particularly attractive
since they are also super-renormalizable. Given that the dynamics of the gauge
field is responsible for the strong interaction and for the formation of bound
states, it comes as no surprise that a great deal of effort has gone into the
investigation of bound states of pure glue in supersymmetric 
models~\cite{lpreview,alp2}.  While such theories capture the essential 
properties of the mass spectrum, and some of them are relevant for the string
theory~\cite{tay98}, the wave functions are quite different from the ones for 
mesons and baryons.  Extensive study of the meson spectrum of non-supersymmetric
theories has been done (see~\cite{bpp98} for a review). Recently an initial
study addressed some of these states in the context of supersymmetric
models~\cite{Lunin:2001im}.  

Throughout this paper we use the word ``meson''
to  indicate the group structure of the state.  Namely we define a meson as a 
bound state whose wave function can be written as a linear combination of 
parton chains, each chain starting and ending with a creation operator in the 
fundamental representation.  In supersymmetric theories, the states defined 
this way can have either bosonic or fermionic statistics.

Previously we saw that the lightest bound states in ${\cal N}=1$
supersymmetric theories are very interesting~\cite{Hiller:2002cu,Hiller:2002pj}.
In supersymmetric Yang--Mills (SYM) theories in 1+1 and 2+1 dimensions,
the lightest bound states in the spectrum are massless 
Bogomol'nyi--Prasad--Sommerfield (BPS) bound states~\cite{lpreview}. These 
states are exactly massless at all couplings.  When we add a Chern--Simons 
(CS) term to the (1+1)-dimensional SYM theory, which gives a mass to the
constituents, we find approximate BPS states. The masses of these states
are approximately independent of the coupling, and at strong coupling these
states are the lightest bound states in the theory~\cite{Hiller:2002cu}.  
In (2+1)-dimensional SYM-CS theory we find that at strong
coupling there are anomalously light bound states~\cite{Hiller:2002pj}. In
both 2+1 and 1+1 dimensions, these interesting states appear because of the 
exact BPS symmetry in the underlying SYM theory.

Here we will look at the lightest bound states of SYM--CS theory with
fundamental matter. The properties of the entire spectrum will be presented
elsewhere~\cite{hpt03}. Again we will see that the lightest bound states are
significantly lighter than one would have naively expected. We will see that
the lightest state is nearly massless compared to the threshold for the bound
states at one unit of adjoint-parton mass.

There are several approaches to the solution of this problem. For QCD-like 
theories, lattice gauge theory is probably the most popular approach since
the lattice approximation
does not break the most important symmetry, gauge symmetry. Similarly for 
supersymmetric theories, supersymmetric discrete light-cone
quantization (SDLCQ)~\cite{bpp98,sakai,lpreview,alp98a} is probably the 
most powerful approach since the discretization does not break the most
important symmetry, supersymmetry. In this paper we consider supersymmetric 
theories and follow this latter approach. To
simplify the calculation we will consider only the large-$N_c$ 
limit~\cite{thooft}, which has proven to be a powerful approximation for 
bound-state calculations. While baryons can be constructed in this 
limit~\cite{WittBar}, they have an infinite number of partons and thus 
practical calculations for such states are complicated.   Note that throughout 
this paper we completely ignore the zero-mode problem~\cite{zm5,thorn};
however, it is clear that considerable progress on this issue
could be made following our earlier work on the zero modes of the 
two-dimensional supersymmetric model with only adjoint fields~\cite{alptzm}.

The paper has the following organization. In Section~\ref{SectDefin} we
consider three-dimensional supersymmetric QCD (SQCD) and dimensionally reduce 
it to 1+1 dimensions. We perform the light-cone quantization of the resulting 
theory by applying canonical commutation relations at fixed 
$x^+\equiv(x^0+x^1)/\sqrt{2}$ and choosing
the light-cone gauge ($A^+=0$) for the vector field. After solving the 
constraint equations, we obtain a model containing 4 dynamical fields. We
construct the supercharge for this dimensionally reduced theory.  In 
Section~\ref{SecMesons} we discuss the structure of the lighter meson 
bound states. In Section~\ref{sec:SuperCS} we discuss the addition of
a Chern-Simons (CS) term to the supercharge~\cite{CSSYM1+1,dunne,witten2}.
We explain that, in the context of this model, this is equivalent to adding a
mass to the partons in the adjoint representation. In particular, in 
Section~\ref{secnum} we discuss bound-state solutions that we find for the
lowest mass states.  We also present an analysis of the convergence in the
longitudinal resolution.  Finally, in this same section we display the 
structure function for the lowest state. In Section~\ref{secdis} we discuss
our results and the future directions that are indicated by this research.

\section{Supersymmetric systems with fundamental matter.}
\label{SectDefin}
We consider the supersymmetric models in two dimensions which
can  be obtained as the result of dimensional reduction of SQCD$_{2+1}$.
Our starting point is the three-dimensional action
\bea\label{action}
S&=&\int d^3x\mbox{tr}\left(-\frac{1}{4}F_{\mu\nu}F^{\mu\nu}+
\frac{i}{2}{\bar\Lambda}
\Gamma^\mu D_\mu \Lambda +D_\mu \xi^\dagger D^\mu \xi+
i{\bar\Psi} D_\mu\Gamma^\mu\Psi\right.\nonumber\\
&&-\left.g\left[{\bar\Psi}\Lambda\xi+
\xi^\dagger{\bar\Lambda}\Psi\right]\right)\,.
\eea
This action describes a system of a gauge field $A_\mu$, representing
gluons, and its superpartner $\Lambda$, representing gluinos, both taking 
values in the adjoint representation of $SU(N_c)$,
and two complex fields, a scalar $\xi$ representing squarks and a Dirac 
fermion $\Psi$ representing quarks, transforming according to the fundamental 
representation of the same group. In matrix notation the covariant
derivatives are given by
\be
D_\mu\Lambda=\partial_\mu\Lambda+ig[A_\mu,\Lambda]\,,\quad
D_\mu\xi=\partial_\mu\xi+igA_\mu\xi\,,\quad
D_\mu\Psi=\partial_\mu\Psi+igA_\mu\Psi\,.
\ee
The action (\ref{action}) is invariant under the following supersymmetry 
transformations, which are parameterized by a two-component Majorana fermion 
$\varepsilon$:
\bea
&&\delta A_\mu=\frac{i}{2}{\bar\varepsilon}\Gamma_\mu\Lambda\,,\qquad
\delta\Lambda=\frac{1}{4}F_{\mu\nu}\Gamma^{\mu\nu}\varepsilon\,,\nonumber\\
&&\delta\xi=\frac{i}{2}{\bar\varepsilon}\Psi\,,\qquad
\delta\Psi=-\frac{1}{2}\Gamma^\mu\varepsilon D_\mu\xi\,.
\eea
Using standard techniques one can construct the Noether current corresponding
to these transformations as
\bea\label{sucurrent}
{\bar\varepsilon}q^\mu&=&\frac{i}{4}{\bar\varepsilon}\Gamma^{\alpha\beta}
\Gamma^\mu\mbox{tr}\left(\Lambda F_{\alpha\beta}\right)+
\frac{i}{2}D^\mu\xi^\dagger\
{\bar\varepsilon}\Psi+\frac{i}{2}\xi^\dagger{\bar\varepsilon}\Gamma^{\mu\nu}
D_\nu\Psi\nonumber\\
&&-\frac{i}{2}{\bar\Psi}\varepsilon D^\mu\xi+\frac{i}{2}D_\nu
{\bar\Psi}\Gamma^{\mu\nu}\varepsilon\xi\,.
\eea

We will consider the reduction of this system to two dimensions, which means
that the field configurations are assumed to be independent of the
transverse coordinate $x^2$. In the resulting two-dimensional system we will
implement light-cone quantization, where the initial conditions as
well as canonical commutation relations are imposed on a light-like
surface $x^+=\mbox{\em const}$. In particular, we construct the supercharge by
integrating the current (\ref{sucurrent}) over the light-like surface to obtain
\bea\label{sucharge}
{\bar\varepsilon}Q&=&\int dx^-dx^2\left(
\frac{i}{4}{\bar\varepsilon}\Gamma^{\alpha\beta}
\Gamma^+\mbox{tr}\left(\Lambda F_{\alpha\beta}\right)+
\frac{i}{2}D_-\xi^\dagger\
{\bar\varepsilon}\Psi+\frac{i}{2}\xi^\dagger{\bar\varepsilon}\Gamma^{+\nu}
D_\nu\Psi\right.\nonumber\\
&&-\left.\frac{i}{2}{\bar\Psi}\varepsilon D^+\xi+\frac{i}{2}D_\nu
{\bar\Psi}\Gamma^{+\nu}\varepsilon\xi\right)\,.
\eea
Since all fields are assumed to be independent of $x^2$, the integration
over this coordinate gives just a constant factor, which we absorb by a field
redefinition. 

If we use the following specific representation for the Dirac matrices in three
dimensions:
\be
\Gamma^0=\sigma_2\,,\qquad \Gamma^1=i\sigma_1\,,\qquad \Gamma^2=i\sigma_3\,,
\ee
the Majorana fermion $\Lambda$ can be chosen to be real.  It is also
convenient to write the fermion fields and the supercharge in component 
form as
\be
\Lambda=\left(\lambda,{\tilde\lambda}\right)^T\,,\qquad
\Psi=\left(\psi,{\tilde\psi}\right)^T\,,\qquad
Q=\left(Q^+,Q^-\right)^T\,.
\ee
In terms of this decomposition the superalgebra has an explicit $(1,1)$ form
\be\label{sualg}
\{Q^+,Q^+\}=2\sqrt{2}P^+\,,\qquad \{Q^-,Q^-\}=2\sqrt{2}P^-\,,\qquad
\{Q^+,Q^-\}=0\,.
\ee
The SDLCQ method exploits this superalgebra by constructing $P^-$
from a discrete approximation to $Q^-$~\cite{sakai}, rather than
directly discretizing $P^-$, as is done in ordinary DLCQ~\cite{bpp98}.
 
To begin to eliminate nondynamical fields, we impose the light-cone gauge
($A^+=0$).  Then the supercharges are given by
\bea\label{Qplus}
Q^+&=&2\int dx^-\left(\lambda\partial_-A^2+
\frac{i}{2}\partial_-\xi^\dagger\psi-\frac{i}{2}\psi^\dagger\partial_-\xi-
\frac{i}{2}\xi^\dagger\partial_-\psi+\frac{i}{2}\partial_-\psi^\dagger\xi
\right)\,,\\
Q^-&=&-2\int dx^-\left(-\lambda\partial_-A^-+
i\xi^\dagger D_2\psi-iD_2\psi^\dagger\xi+\frac{i}{\sqrt{2}}
\partial_-({\tilde\psi}^\dagger\xi-\xi^\dagger{\tilde\psi})\right)\,.\nonumber
\\
\eea
Note that apart from  a total derivative these expressions involve only
left-moving components of the fermions ($\lambda$ and $\psi$). In fact, in the
light-cone formulation only these components are dynamical. To see this we
consider the equations of
motion that follow from the action (\ref{action}), in light-cone gauge.
Three of them serve as constraints rather than as dynamical statements;
they are
\bea
\partial_-{\tilde\lambda}&=&-\frac{ig}{\sqrt{2}}
\left([A^2,\lambda]+i\xi\psi^\dagger-i\psi\xi^\dagger\right),\\
\partial_-{\tilde\psi}&=&-\frac{ig}{\sqrt{2}}A^2\psi+
\frac{g}{\sqrt{2}}\lambda\xi\,,
\eea 
and
\be\label{constraint}
\partial^2_-A^-=gJ\,,
\ee
with
\be
J\equiv i[A^2,\partial_-A^2]+
\frac{1}{\sqrt{2}}\{\lambda,\lambda\}-ih\partial_-\xi\xi^\dagger+
i\xi\partial_-\xi^\dagger+\sqrt{2}\psi\psi^\dagger\,.
\ee
Apart from the zero-mode problem~\cite{zm5}, one can invert the last
constraint to write the auxiliary field $A^-$ in terms of physical degrees
of freedom. Substituting the result into the expression for the supercharge 
and omitting the boundary term, we get
\be\label{Qminus}
Q^-= Q^-_s +Q^-_1 +Q^-_2 +Q^-_3\,.
\ee
where $Q^-_s$ is the supercharge of pure adjoint matter~\cite{sakai} and
\bea
Q^-_1&=&-\frac{g}{\sqrt{2}}\int dx^-\left(
i\sqrt{2}\xi\partial_-\xi^\dagger-i\sqrt{2}\partial_-\xi\xi^\dagger\right)
\frac{1}{\partial_-}\lambda\,, \\
Q^-_2&=&-\frac{g}{\sqrt{2}}\int dx^-\left(
2\psi\psi^\dagger\right)
\frac{1}{\partial_-}\lambda\,, \\
Q^-_3&=&-2g\int dx^-\left(\xi^\dagger A^2\psi+\psi^\dagger A^2\xi\right)\,.
\eea

In order to solve the bound-state problem $2P^+P^-|M\rangle=M^2|M\rangle$,
we apply the methods of
SDLCQ. Namely we compactify the two-dimensional theory on a
light-like circle ($-L<x^-<L$), and impose periodic boundary conditions on
all physical fields. This leads to the following mode expansions:
\bea
A^2_{ij}(0,x^-)&=&\frac{1}{\sqrt{4\pi}}\sum_{k=1}^{\infty}\frac{1}{\sqrt{k}}
\left(a_{ij}(k)e^{-ik\pi x^-/L}+a^\dagger_{ji}(k)e^{ik\pi x^-/L}\right)\,,\\
\label{expandLambda}
\lambda_{ij}(0,x^-)&=&\frac{1}{2^{\frac{1}{4}}\sqrt{2L}}\sum_{k=1}^{\infty}
\left(b_{ij}(k)e^{-ik\pi x^-/L}+b^\dagger_{ji}(k)e^{ik\pi x^-/L}\right)\,,\\
\xi_i(0,x^-)&=&\frac{1}{\sqrt{4\pi}}\sum_{k=1}^{\infty}\frac{1}{\sqrt{k}}
\left(c_i(k)e^{-ik\pi x^-/L}+{\tilde c}^\dagger_{i}(k)e^{ik\pi
x^-/L}\right)\,,\\
\label{expandPsi}
\psi_{i}(0,x^-)&=&\frac{1}{2^{\frac{1}{4}}\sqrt{2L}}\sum_{k=1}^{\infty}
\left(d_{i}(k)e^{-ik\pi x^-/L}+{\tilde d}^\dagger_{i}(k)e^{ik\pi
x^-/L}\right)\,.
\eea 
We drop the zero modes of the fields; including them
could lead to new and interesting effects (see~\cite{alptzm}, for example),
but this is beyond the scope of this work.
In the light-cone formalism one treats $x^+$ as the time direction, thus
the commutation relations between fields and their momenta are imposed on
the surface $x^+=0$.  For the system under consideration this means that
\bea\label{CanComRelField}
\left[A_{ij}^2(0,x^-),\partial_-A_{kl}^2(0,y^-)\right]&=&
i\left(\delta_{il}\delta_{kj}-\frac{1}{N}\delta_{ij}\delta_{kl}\right)
\delta(x^--y^-)\,,\\
\left\{\lambda_{ij}(0,x^-),\lambda_{kl}(0,y^-)\right\}&=&\sqrt{2}
\left(\delta_{il}\delta_{kj}-\frac{1}{N}\delta_{ij}\delta_{kl}\right)
\delta(x^--y^-)\,,\\
\left[\xi_i(0,x^-),\partial_-\xi_j(0,y^-)\right]&=&
i\delta_{ij}\delta(x^--y^-)\,,\\
\left\{\psi_{i}(0,x^-),\psi_{j}(0,y^-)\right\}&=&\sqrt{2}
\delta_{ij}\delta(x^--y^-)\,.
\eea
These relations can be rewritten in terms of creation and annihilation
operators as
\be
\left[a_{ij},a^\dagger_{kl}\right]=
\left(\delta_{il}\delta_{kj}-\frac{1}{N}\delta_{ij}\delta_{kl}\right)\,,\quad
\left\{b_{ij},b^\dagger_{kl}\right\}=
\left(\delta_{il}\delta_{kj}-\frac{1}{N}\delta_{ij}\delta_{kl}\right)\,,
\ee
\be
\left[c_{i},c^\dagger_{j}\right]=\delta_{ij}\,,\quad
\left[{\tilde c}_{i},{\tilde c}^\dagger_{j}\right]=\delta_{ij}\,,\quad
\left\{d_{i},{d}^\dagger_{j}\right\}=\delta_{ij} \quad
\left\{{\tilde d}_{i},{\tilde d}^\dagger_{j}\right\}=\delta_{ij}\,.
\ee
In this paper we will discuss numerical results obtained
in the large-$N_c$ limit, i.e.\ we neglect $1/N_c$ terms in the above
expressions. Although $1/N_c$ corrections may have interesting
consequences, they are beyond the scope of this work.

\section{Mesons}
\label{SecMesons}
We will consider here only meson-like states. In the large-$N_c$ approximation 
these are color-singlet states with exactly two partons in the fundamental
representation.  The boson bound states will have either two bosons in the 
fundamental representation or two fermions in the fundamental representation. 
In general a boson bound state will have a combination of these types of 
contributions.  Because this theory and the numerical formalism are exactly
supersymmetric, for each boson bound state there will be a degenerate bound
state that is a fermion. The fermion bound state 
will have one fermion in the fundamental
representation and one boson in the fundamental representation. In the string 
interpretation of these theories, such states would correspond 
to open strings with freely moving endpoints. In the language of QCD, the 
model corresponds to a system of interacting gluons and gluinos which bind
dynamical (s)quarks and anti-(s)quarks. In the large-$N_c$ limit we will have
to consider only a single (s)quark--anti-(s)quark pair.  Thus the Fock space
is constructed from states of the following type:
\be\label{state}
{\bar f}^\dagger_{i_1}(k_1) a^\dagger_{i_1i_2}(k_2)\dots
b^\dagger_{i_ni_{n+1}}(k_{n-1})\dots f^\dagger_{i_p}(k_n)|0\rangle\,.
\ee
Here $\bar{f}_i^\dagger$ and $f_i^\dagger$ each create one of the 
fundamental partons, and $|0\rangle$ is the vacuum annihilated by $a_{ij}$,
$b_{ij}$, $c_i$, ${\tilde c}_i$, $d_i$, and ${\tilde d}_i$.  In this basis 
$P^+$ is automatically diagonal.

The three supercharges that govern the behavior of the
fundamental matter in these states are $Q^-_1$, $Q^-_2$, and $Q^-_3$.
For example, after substituting the expansions (\ref{expandLambda}) and
(\ref{expandPsi}) one gets the mode decomposition of $Q^-_2$

\begin{eqnarray}
Q^-_2&=&\frac{i2^{-1/4}g\sqrt{L}}{\pi}\sum_{k_1,k_2=1}^\infty
 \frac{1}{k_1}[{\tilde d}^\dagger_i(k_2)
 b^\dagger_{ij}(k_1){\tilde d}_j(k_1+k_2)+
 {\tilde d}^\dagger_j(k_1{+}k_2){\tilde d}_i(k_2)b_{ij}(k_1) 
\nonumber \\
&+&d^\dagger_i(k_2)b^\dagger_{ij}(k_1) d_j(k_1+k_2)+
d^\dagger_j(k_1+k_2) d_i(k_2)b_{ij}(k_1) ]\,.
\label{FrmOnlSuchMode}
\end{eqnarray}

The other color-singlet bound states in this theory are states that are
composed of traces of only adjoint mesons. These can be thought of as loops. 
At finite $N_c$ this theory has interactions that break these loops and 
insert a pair of fundamental partons, making an open-string state. This type
of interaction can of course also form loops from open strings and break open
strings into two. In principle, a calculation of the spectrum of such a 
finite-$N_c$ theory is within the reach of SDLCQ. The only significant change 
is to include states in the basis with more than one color trace.

\section{Supersymmetric Chern--Simons theory} \label{sec:SuperCS}

The CS term we use in this calculation is obtained by starting with a 
CS term in 2+1 dimensions and reducing it to 1+1 dimensions.  This
term has the effect of adding a mass for the adjoint partons.
In this calculation we are including fundamental matter because we are 
interested in QCD-like meson bound states. Without a mass for the adjoint 
matter, this theory is known to produce very
long light chains of adjoint partons. In a finite-$N_c$ calculation we 
would not have these very long chains because they would break, but
in the large-$N_c$ approximation they do not. While SDLCQ can be used to do
finite-$N_c$ calculations, it is much easier to add a mass to restrict
the number of adjoint partons in our bound states.  We choose the CS
mechanism to give the adjoint partons a mass because we can do this 
without breaking the supersymmetry.

The Lagrangian  of this theory is
\begin{equation} \label{Lagrangian}
{\cal L}={\cal L}_{\rm SQCD}+\frac{\kappa}{2}{\cal L}_{\rm CS}\,,
\end{equation}
where ${\cal L}_{\rm SQCD}$ is the SQCD Lagrangian we discussed earlier,
$\kappa$ is the CS coupling, and
\begin{equation}
{\cal L}_{\rm CS}=\epsilon^{\mu\nu\lambda}\left(A_{\mu}
\partial_{\nu}A_{\lambda}+\frac{2i}{3}gA_\mu A_\nu A_\lambda \right)
+2\bar{\Psi}\Psi\,. \label{eq:CSLagrangian}
\end{equation}
A trace of the color matrices is understood.
The discrete version of the CS part of the supercharge in 1+1
dimensions is
\begin{equation} \label{qcs}
Q^-_{CS}=\left(\frac{2^{-1/4}\sqrt{L}}{\sqrt{\pi}}\right)
\sum_{n}\frac{\kappa}{\sqrt{n}}
\left(A^{\dagger}(n)B(n)+B^{\dagger}(n)A(n)\right)\,,
\end{equation}
where $A$ and $B$ are rescaled discrete field operators
\begin{equation}
A(n)\equiv \sqrt{\frac{\pi}{L}}a_{ij}(n\pi/L)\,,  \quad\quad
B(n)\equiv \sqrt{\frac{\pi}{L}}b_{ij}(n\pi/L)\,.
\end{equation}
It is important to compare $Q_{\rm CS}^-$ with the supercharge
for ${\cal N}=1$ SYM in 2+1 dimensions~\cite{hpt2001}, which
has a contribution of the form
\begin{equation}
Q^-_{\perp}=i\left(\frac{g2^{-1/4}\sqrt{L}}{\sqrt{\pi}}\right)
                \sum_{n,n_\perp}\frac{k_\perp}{\sqrt{n}}
                     \left(A^{\dagger}(n,n_\perp)B(n,n_\perp)
                         -B^{\dagger}(n,n_\perp)A(n,n_\perp)\right)\,,
\end{equation}
where $k_\perp=2\pi n_\perp/L_\perp$ is the discrete transverse 
momentum.  Notice that $k_\perp$ and $\kappa$ enter the supercharge in
very similar ways.  Because the light-cone energy is of the form
$(k^2_\perp + m^2)/k^+$, $k_\perp$ behaves like a
mass, and therefore $\kappa$ also behaves in many ways like a 
mass for the adjoint particles.

The partons in the fundamental representation in this theory will remain
massless. Of course, in a more physical theory the supersymmetry would be
badly broken; the squark would acquire a large mass, and only the quarks
would remain nearly massless.

\section{Numerical results}
\label{secnum}
This SYM-CS theory with fundamental matter has two dimensionful parameters 
with dimension of a mass squared, the YM coupling squared $g^2 N_c/\pi$
and the CS coupling squared $\kappa^2$.  The latter is also the mass squared
of the partons in the adjoint representation.  Furthermore, we are only 
considering meson bound states. These are states of the form shown in 
Eq.~(\ref{state}) with
two fundamental partons linked by partons in the adjoint representation. Since 
we are working in the large-$N_c$ approximation, this class of states is
disconnected from the other allowed class of pure adjoint matter
bound states and multiparticle states.

This theory also has a $Z_2$ symmetry~\cite{kutasov93}
which is very useful in labeling the states and reducing the dimension of 
the Fock basis that one has to consider in any one diagonalization step.
For this theory the $Z_2$ symmetry divides the basis into
states with an even or odd number of gluons. Here we will focus on the lowest
mass states in the sector with an odd number of adjoint bosons. The
lowest mass state in the sector with an even number of bosons has a mass that
converges to about $0.2\kappa^2$; at strong coupling this state converges
significantly slower than the state we consider here, and it will be
presented in detail in a future publication, where we will discuss the entire
spectrum of this theory~\cite{hpt03}.

\begin{figure}
\centerline{
\psfig{figure=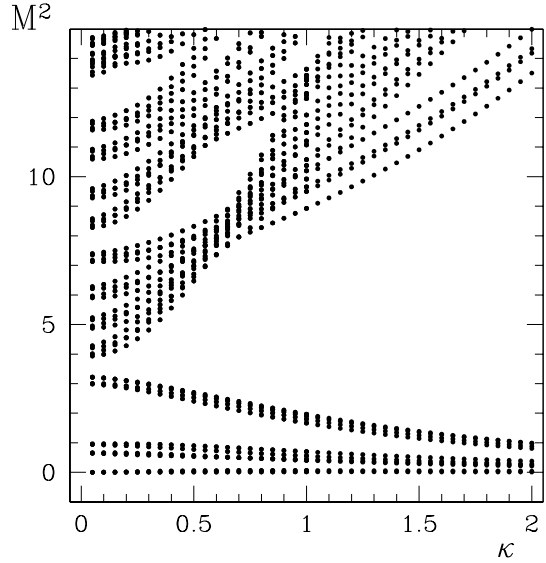,width=7.5cm,angle=0}}
\caption{The mass-squared spectrum, in units of $g^2N_c/\pi$,
as a function of $\kappa$, in units of $g\sqrt{N_c/\pi}$,
at a resolution of $K=6$.}
\label{fig:spectrum}
\end{figure}

We find that the $Z_2$-odd spectrum of meson bound states divides into 
two bands of states, a very light band and a heavy band, as can be seen
in Fig.~\ref{fig:spectrum}. This is easy to understand if we
start by considering large $\kappa$. At large $\kappa$, the light band is
primarily composed of two fundamental partons and a small mixture of adjoint 
partons, and the heavy band is composed of bound states that have at
least one adjoint parton. In the limit of very large $\kappa$, all the particles
in the low-mass band become massless. Here we will focus on the low-mass band
but keep $\kappa$ at or below $g\sqrt{N_c/\pi}$. These moderate values
of $\kappa$ will allow a significant mixture of
adjoint matter in the bound states. The lowest mass state at 
$\kappa=g\sqrt{N_c/\pi}$ has an average particle count of about three, 
two fundamental partons and one adjoint parton.

If this state has on average one adjoint parton, which has mass $\kappa^2$,
and the fundamental partons are massless, one would expect the
lowest bound state in the spectrum to have a mass of order $\kappa^2$. We
actually see that the lowest mass state is nearly massless.  We appear to
have found a state similar to those found in ${\cal N}=1$ SYM-CS
theory~\cite{Hiller:2002cu,Hiller:2002pj}. The lowest
mass state is anomalously light and in fact nearly massless.

\begin{figure}
\begin{tabular}{cc}
\psfig{figure=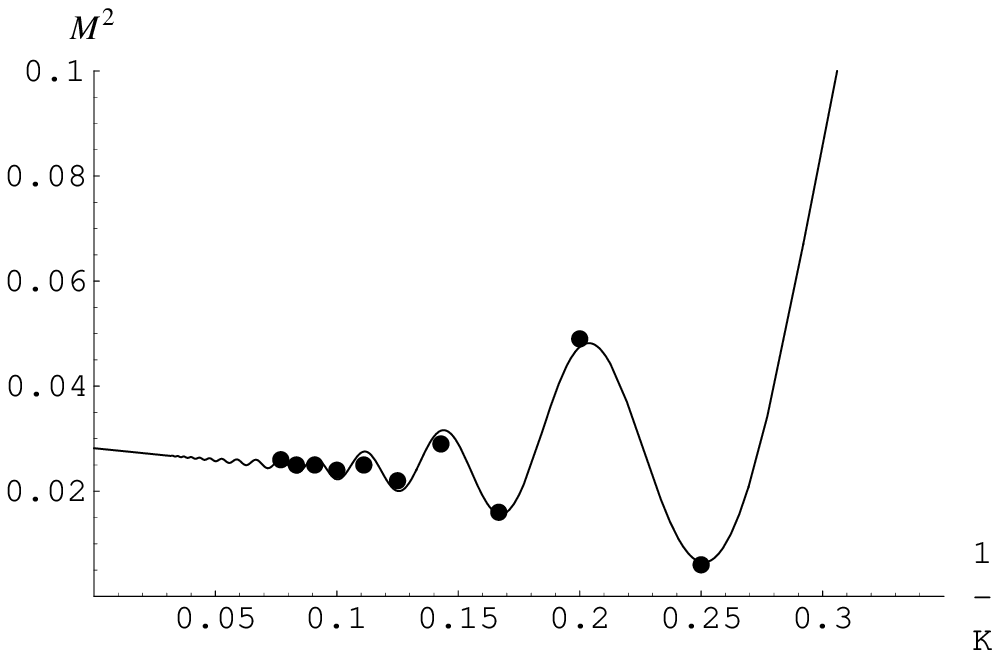,width=7.5cm,angle=0}&
\psfig{figure=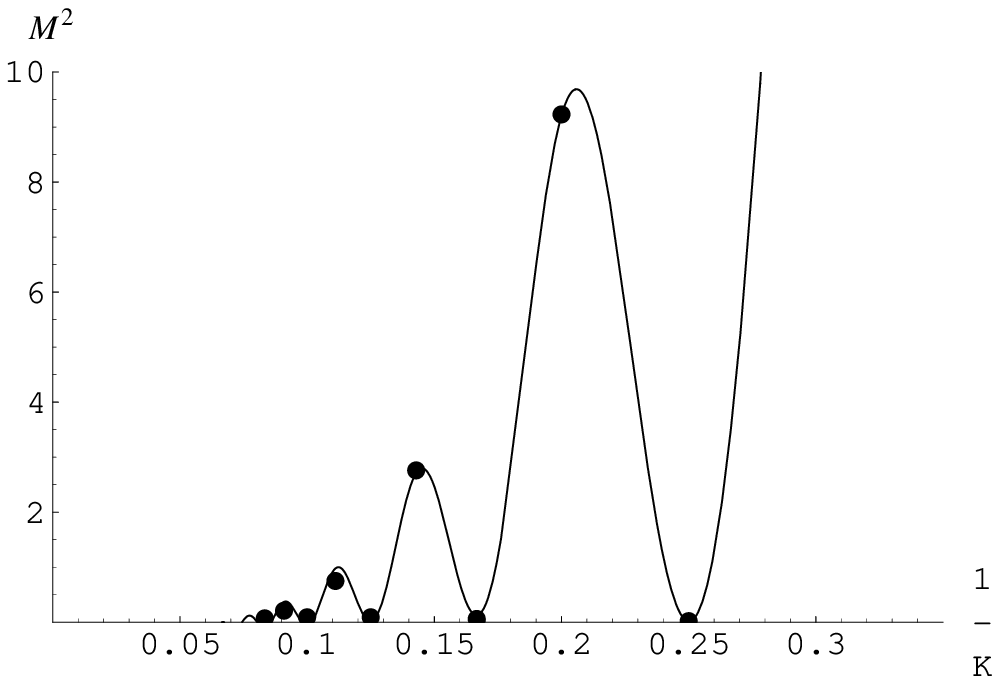,width=7.5cm,angle=0}\\
(a) & (b)
\end{tabular}
\caption{The mass squared of the lowest mass state, in units of
$g^2N_c/\pi$, as a function of $1/K$
for (a) $\kappa=g\sqrt{N_c/\pi}$ and (b) $\kappa=0.1\,g\sqrt{N_c/\pi}$.
The solid curve is a fit to the computed points.}
\label{da}
\end{figure}

The mass of the lightest bound state as a function of the resolution
is shown in Fig.~\ref{da} for $\kappa=g\sqrt{N_c/\pi}$ and
$\kappa=0.1\,g\sqrt{N_c/\pi}$. The convergence plot
shows a rather unusual oscillatory behavior as a function of the resolution
$K$. This type of behavior was seen in a DLCQ study of
(1+1)-dimensional large-$N_c$ QCD coupled to a massive adjoint Majorana 
fermion~\cite{Gross:1997mx}, and the explanation there is that the 
spectrum of two free particles as a function
of the resolution oscillates and therefore a bound state that is in some way
closely related to a free-particle spectrum might oscillate. The
oscillations we see here are in fact much larger than those seen 
in~\cite{Gross:1997mx}. As we discussed above, the low-mass band
in the spectrum is strongly connected to the free spectrum of two fundamental
partons, particularly at large $\kappa$, so some oscillation might be
expected.

The oscillatory behavior made this calculation particularly challenging
numerically. We were forced to go to very high resolution, $K=13$, to be
certain that the spectrum really converged. This was made even more
difficult because we had 4 species of particles in the problem. To be
certain that we were identifying the states properly and isolating the
correct state, it was very useful to calculate a number of properties of 
the states.  These are shown in Table~\ref{table1}. There are, of course,
degenerate fermion bound states whose properties we do not show. We find
that the average number of partons in this state is approximately three 
and that for the most part the adjoint particle is a
boson. We also see that about $2/3$ of the wave function of this state 
is composed of two fundamental bosons and an adjoint parton and about 
$1/3$ of the wave function is made of two fundamental fermions and an 
adjoint parton. Within the context of the standard model, this state is 
primarily a bound state of two squarks and a gluon.
\begin{table}
\center{
\begin{tabular}{|c|c|c|c|c|c|c|}
\hline
K & $M^2$ &$<n>$  &$<n_{aB}>$&$<n_{fB}>$&$<n_{aF}>$&$<n_{fF}>$\\
\hline
3    & 0.178    &    2.30     &    0.30    & 1.03    & 0.01    & 0.97\\
4    & 0.006    &    2.56     & 0.51    & 1.86    & 0.05    & 0.14\\
5    & 0.049    &    2.69     &    0.63    & 1.29    & 0.06    & 0.71\\
6    & 0.016    &    2.83        &    0.75    & 1.71    & 0.08    & 0.30\\
7    & 0.029    &    2.84     &    0.76    & 1.45    & 0.08    & 0.55\\
8    & 0.022 &    2.92     &    0.83    & 1.58    & 0.09    & 0.42\\
9    & 0.025    &    2.92     &    0.83    & 1.49    & 0.10    & 0.51\\
10    & 0.024    &    2.96    &    0.86    & 1.52    & 0.10    & 0.48\\
11    & 0.025    &    2.97    &    0.87    & 1.49    & 0.11    & 0.51\\
12    & 0.025    & 3.00    &    0.89    & 1.48    & 0.11    & 0.52\\
13    & 0.026    &    3.01    &    0.90    & 1.47    & 0.11    & 0.53\\
\hline
\end{tabular}
}
\caption{Properties of the lowest mass boson bound state, including the
average numbers of adjoint bosons $aB$, adjoint fermions $aF$,
fundamental bosons $fB$, and fundamental fermions $fF$, for different
values of the resolution $K$.  The mass squared $M^2$ is given in 
units of $g^2N_c/\pi$.  The CS coupling is $\kappa=g\sqrt{N_c/\pi}$.}
\label{table1}
\end{table}

This state is different than the special state that we saw in pure SYM and
SYM-CS theories. Those states had masses that were above threshold. The
state that we are considering here has a mass near zero, and threshold is
at $\kappa^2$.  Thus this is a deeply bound state. In a more
realistic theory, the squark would be very heavy. It is
conceivable that this mechanism would give a binding well 
below the threshold even then.

\begin{figure}
\begin{tabular}{cc}
\psfig{figure=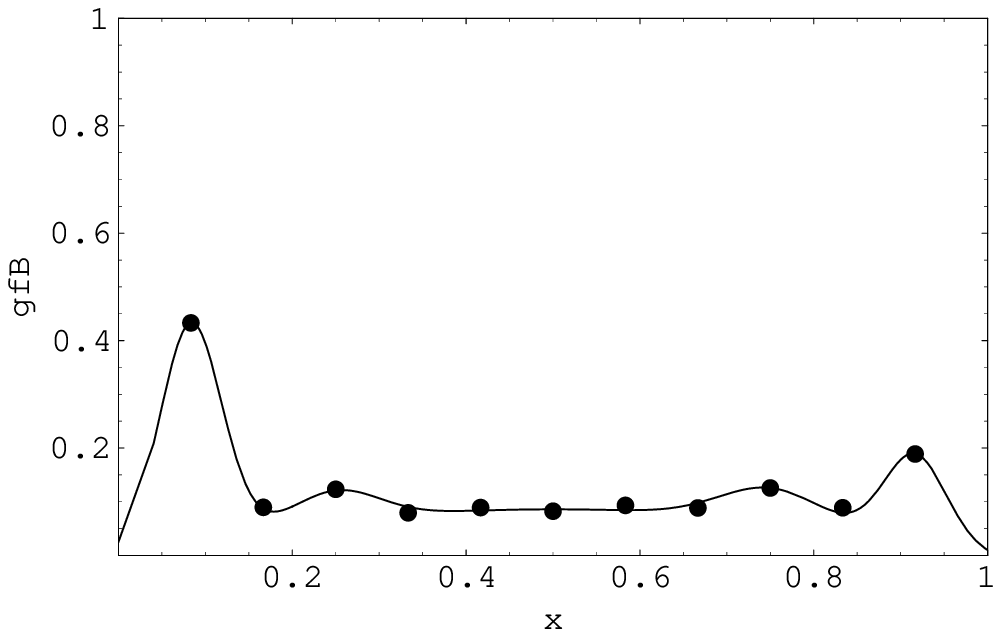,width=7.5cm,angle=0}&
\psfig{figure=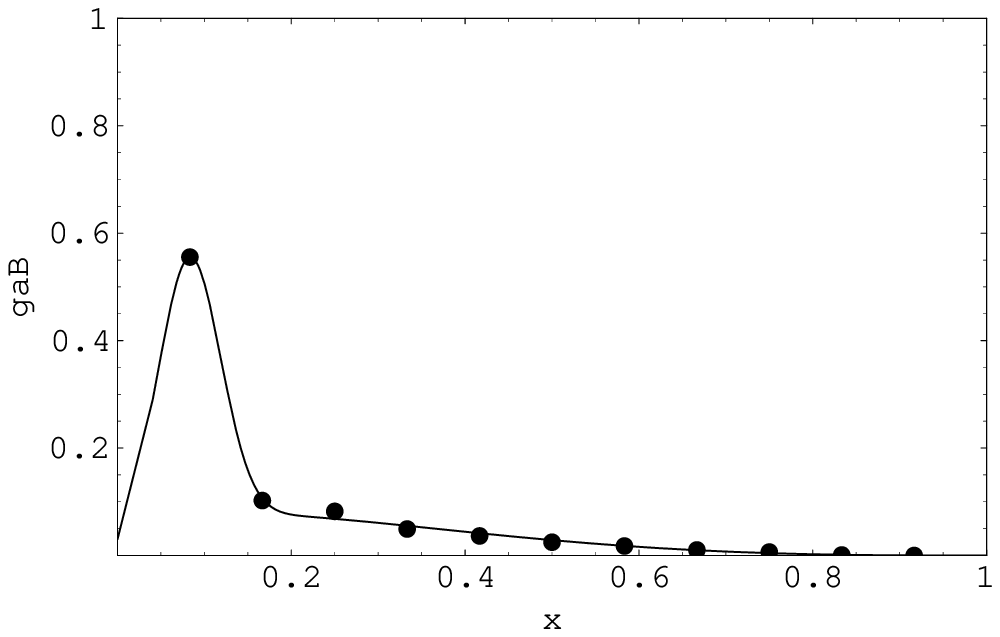,width=7.5cm,angle=0}\\
(a) & (b)
\end{tabular}
\caption{Structure functions of the lowest mass bound state at
resolution $K=12$ for (a) fundamental bosons $fB$ and 
(b) adjoint bosons $aB$, with the CS coupling fixed at
$\kappa=g\sqrt{N_c/\pi}$.  The solid curve is a fit to
the computed points.}
\label{struc}
\end{figure}

It is very instructive to look at the structure functions of this
bound state. We use a standard definition of the structure functions
\begin{eqnarray}
\hat{g}_A(x)&=&\sum_q\int_0^1 dx_1\cdots dx_q
\delta\left(\sum_{i=1}^q x_i-1\right)
\sum_{l=1}^q \delta(x_l-x)\delta^A_{A_l}
|\psi(x_1,\ldots x_q)|^2\,.
\end{eqnarray}
Here $A$ stands for the chosen representation (fundamental or adjoint)
and statistics (boson or fermion) of the constituent of interest.
The sum runs over all parton numbers $q$, and the Kronecker delta
$\delta^A_{A_l}$ selects partons with matching representation
and statistics $A_l$.  The discrete approximation $g_A$ to the structure 
function $\hat{g}_A$ with harmonic resolution $K$ is
\begin{eqnarray}
{g}_A(n)&=&\sum_{q=2}^K\sum_{n_1,\ldots,n_q=1}^{K-q}
\delta\left(\sum_{i=1}^q n_i-K\right)
\sum_{l=1}^q \delta^{n_l}_n\delta^A_{A_l}
|\psi(n_1,\ldots n_q)|^2\,.
\end{eqnarray}
The functions $g_{A}(n)$ are normalized so that summation over the
argument gives the average  number of type A particles;
their sum is then the average parton
number, and we compute these sums as a check. We plot the structure
functions as functions of the longitudinal momentum fraction
$x=k/P^+=n/K$ carried by an individual parton.  In this lightest state the
bulk of the partons are fundamental or adjoint bosons, and their structure
functions are shown in Fig.~\ref{struc}. In the fit shown we have forced the
fitting function to vanish at $x=0$. This is an assumption, and a fitting
function which goes to a finite value at $x=0$ would also work.  We see 
that both of the distributions in Fig.~\ref{struc} are peaked at small
$x$. This reflects strong binding of the fundamental partons, allowing them 
to be widely separated in momentum, combined with only a small contribution 
to the momentum from the adjoint boson.  We cannot, however, fit this peak 
at small $x$ with a divergent function, such as $1/x$ or $\log x$, since such
a structure function would imply that the average number of partons
also diverges.  As we probe smaller and smaller $x$ by increasing
the resolution, we would see this divergence.  In Table~\ref{table1} we see
instead that the average number of partons converges at high resolution to 
about three. Such strong peaks at low momentum are not unique to
this state, and we will discuss this further when we consider the full
spectrum~\cite{hpt03}.

\section{Discussion.}
\label{secdis}
In this work we studied the lowest mass meson of SYM-CS theory with
fundamental matter in 1+1 dimensions. The CS term is included to give
masses to the adjoint partons.  The calculations were performed at large
$N_c$ in the framework of SDLCQ; namely, we compactified the light-like
coordinate $x^-$ on a finite circle and calculated the Hamiltonian as the
square of a supercharge $Q^-$, which we then diagonalized numerically.

In previous work we have found that the lowest mass states of a number of
${\cal N}=1$ supersymmetric theories solved at large $N_c$ using SDLCQ have
very interesting properties. In SYM in 1+1 and 2+1 dimensions we found
that there are a number of exactly massless BPS states. In SYM-CS theory
in 1+1 dimensions we saw that at strong coupling the lightest states are
approximately BPS states whose masses are independent of the YM coupling. 
In SYM-CS theories in 2+1 dimensions at strong coupling there is again an
anomalously light bound state.

Now in this SYM-CS theory with fundamental matter in 1+1 dimensions we 
find a very light bound state composed primarily of squarks and gluons,
and we find that this persists at intermediate and strong coupling. This 
state is nearly massless and well below the threshold for
the spectrum. The structure function of this bound state shows that the 
dynamics of this theory tend to maximize the number of 
small-longitudinal-momentum partons in the bound states.
From the SDLCQ numerical perspective the states are interesting because of the
oscillatory convergence in the resolution. To be certain of this convergence
we pushed our resolution to $K=13$, the highest we have attained in a 
(1+1)-dimensional problem without truncating the basis. A study of the 
entire spectrum of this theory will be presented elsewhere~\cite{hpt03}.

There remains a considerable amount of work to be done on SYM theories with
fundamental matter. The most straightforward extension of the present work
is to consider calculations in 2+1
dimensions~\cite{Antonuccio:1999zu,Haney:2000tk,hpt2001}.  The ${\cal N}=1$
theory in 2+1 dimensions is easily within our reach. Beyond that the  
${\cal N}=2$ theory~\cite{Antonuccio:1998mq} in 2+1 dimensions, which is 
the dimensional reduction of the ${\cal N}=1$ theory 3+1 dimensions, will be 
very interesting.

\section*{Acknowledgments}

This work was supported in part by the U.S. Department of Energy.
One of us (S.P.) would like to acknowledge the Aspen Center for Physics 
where part of this work was completed.

\end{document}